\documentclass[prl,twocolumn,showpacs,preprintnumbers,amsmath,amssymb]{revtex4}
\usepackage{graphicx}
\usepackage{dcolumn}
\usepackage{bm}
\newcommand{\Eq}[1]{Eq.~({\protect\ref{#1}})}

\newcommand{\Fig}[1]{Fig.~\protect\ref{#1}}

\newlength{\Tatescale}
\newlength{\figwidth}
 \newcommand{\Cut}[1]{}
\begin{document}
\title{
Nuclear Force from Lattice QCD
}
\author{
  N.~Ishii$^{1,2}$, S.~Aoki$^{3,4}$ and T.~Hatsuda$^2$
}
\affiliation{$^1$
  Center for Computational Sciences,
  University of Tsukuba,
  Tsukuba 305--8577, Ibaraki, JAPAN,
}
\affiliation{$^2$
  Department of Physics,
  University of Tokyo,
  Tokyo 113--0033, JAPAN,
}
\affiliation{$^3$
  Graduate School of Pure and Applied Sciences,
  University of Tsukuba,
  Tsukuba 305--8571, Ibaraki, JAPAN
}
\affiliation{$^4$
 RIKEN BNL Research Center, 
 Brookhaven National Laboratory, Upton, 
 New York 11973, USA
}
%\date{\bf\today}
 
\begin{abstract}
 Nucleon-nucleon (NN) potential is  studied by lattice QCD simulations
  in the quenched approximation,  using the plaquette gauge action and
  the  Wilson  quark action  on  a  $32^4$  ($\simeq $  (4.4  fm)$^4$)
  lattice.   {A  NN  potential  $V_{\rm  NN}(r)$ is  defined  from  the
  equal-time  Bethe-Salpeter  amplitude  with  a  local  interpolating
  operator for  the nucleon.   By studying the  NN interaction  in the
  $^1$S$_0$ and $^3$S$_1$  channels, we show that the  central part of
  $V_{\rm NN}(r)$ has a strong repulsive  core of a few hundred MeV at
  short distances ($r  \alt 0.5$ fm) surrounded by  an attractive well
  at medium  and long distances.}  These features  are consistent with
  the { known phenomenological features} of the nuclear force.
\end{abstract}

\maketitle
%%%%%%%%%%%%%%%%%%%%%%%%%%%%%%%%%%%%%%%%%%%%%%%%%%%%%%%%%%%%%%%%%%%%%%%%%%%%%%%
% The main text
%%%%%%%%%%%%%%%%%%%%%%%%%%%%%%%%%%%%%%%%%%%%%%%%%%%%%%%%%%%%%%%%%%%%%%%%%%%%%%%
 
  More than  70 years ago, Yukawa  introduced the pion  to account for
  the  strong interaction  between  the nucleons  (the nuclear  force)
  \cite{yukawa}.  Since then,  enormous efforts  have been  devoted to
  understand {  the nucleon-nucleon (NN) interaction}  at low energies
  both from theoretical and experimental points of view.
\begin{figure}[b]
\begin{center}
\includegraphics[width=0.7\figwidth]{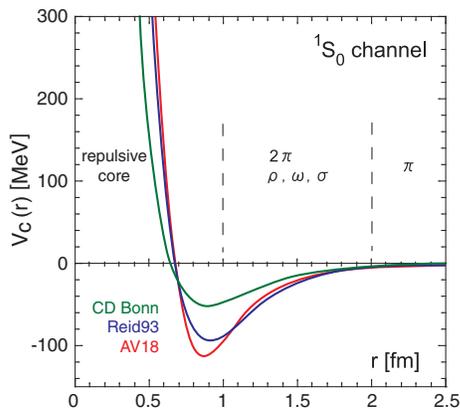}
\end{center}
\caption{Three examples of the modern  
NN potential in the $^1$S$_0$ (spin singlet and s-wave)
channel: CD-Bonn \cite{Machleidt:2000ge},
 Reid93 \cite{Stoks:1994wp} and 
 AV18 \cite{Wiringa:1994wb} from the top at 
 $r=0.8$ fm.}
\label{phen-pot}
\end{figure} 
  As  shown in Fig.\ref{phen-pot},  { phenomenological  NN potentials}
are   thought  to   be   characterized  by   three  distinct   regions
\cite{NN-review-1,NN-review-2}: The long range part ($r \agt 2$ fm) is
well understood  and is  dominated by the  { one pion  exchange}.  The
medium  range  part  ($1\  {\rm  fm}  \alt  r  \alt  2$  fm)  receives
significant contributions  from the exchange of  multi-pions and heavy
mesons  ($\rho$, $\omega$, and  $\sigma$).  The  short range  part ($r
\alt  1$  fm) is  empirically  known  to  have strong  repulsive  core
\cite{jastrow}, which is  essential { not only} for  describing the NN
scattering  data, {  but also}  for  the stability  and saturation  of
atomic nuclei, for determining the  maximum mass of neutron stars, and
for igniting the Type II supernova explosions \cite{VJ}.  Although the
origin  of  the  repulsive  core   must  be  closely  related  to  the
quark-gluon  structure of the  nucleon, it  has been  {a long-standing
open questions in QCD} \cite{oka}.

 In this Letter, we report our  first { serious attempt} to attack the
 {problem   of   nuclear   force   from}   lattice   QCD   simulations
 \cite{IAH-lat06}.  The essential idea  is to {define} a NN potential
 from the  equal-time Bethe-Salpeter (BS) {amplitude of  the two local
 interpolating operators separated by distance $r$} \cite{note1}.
%  It satisfies the effective Schr\"{o}dinger equation
% in the  non-relativistic regime. 
  This type of BS amplitude has been employed by CP-PACS 
 collaboration to study the  $\pi\pi$ scattering on the 
lattice \cite{ishizuka}.
 As we  shall  see  below,  our NN potential shows a
 strong repulsive core  of about a few hundred MeV at short distances
 surrounded by an attraction at medium and long distances
 in the s-wave channel.

 Let us start with an effective
 Schr\"{o}dinger equation  obtained  from the BS amplitude 
  for two nucleons at low energies \cite{ishizuka,IAH_full}:
\begin{equation}
  -\frac1{2\mu} \nabla^2 \phi(\vec r)
  + \int d^3 r'\; U(\vec r,{\vec r}')  \phi({\vec r}')
  = E \phi(\vec r),
  \label{schrodinger.eq2}
\end{equation}
where $\mu\equiv m_{\rm N}/2$ and $E$ is the reduced mass of  the nucleon
 and the non-relativistic energy, respectively.
For  the NN scattering at low energies,
 the non-local potential $U$ is represented   as
 $U(\vec r, \vec r')=V_{\rm NN}({\vec r}, \nabla ) \delta ({\vec r}-{\vec r}')$
 with the derivative expansion \cite{NN-review-1}: 
%\begin{eqnarray}
%\! \! \!  \! \! \! 
$V_{\rm NN}
  =   V_{\rm C}(r)
    + V_{\rm T}(r) S_{12}
    + V_{\rm LS}(r) {\vec L}\cdot {\vec S}
    + O(\nabla^2).$
%    
%\label{NN-pot}
%\end{eqnarray}
Here  $S_{12} =
  3 ({\vec \sigma}_1 \cdot \hat  r) ({\vec \sigma}_2 \cdot \hat r)
   - {\vec \sigma}_1 \cdot {\vec \sigma}_2$ is the tensor operator
    with  $\hat r \equiv  |\vec r|/r$,
  ${\vec  S}$   the    total    spin    operator, and 
${\vec L}\equiv -i \vec r \times \vec \nabla$ the relative angular
momentum  operator.  The central NN potential $V_{\rm  C}(r)$,
the tensor potential $V_{\rm T}(r)$
and the spin-orbit potential $V_{\rm LS}(r)$
 can be further decomposed into various spin-isospin channels, e.g.
$V_{\rm C}(r) = V_{\rm C}^{1}(r)
+ V_{\rm C}^{\sigma}(r) {\vec \sigma}_1 \cdot {\vec \sigma}_2
+ V_{\rm C}^{\tau}(r) {\vec \tau}_1 \cdot {\vec \tau}_2
+ V_{\rm C}^{\sigma \tau}(r)
({\vec \sigma}_1 \cdot {\vec \sigma}_2)({\vec \tau}_1 \cdot {\vec \tau}_2)$.
%and $V_i(r) = V_i^{1}(r)
%+ V_{\rm C}^{\tau}(r) {\vec \tau}_1 \cdot {\vec \tau}_2$ ($i={\rm T}, {\rm LS}$).
  In the phenomenological analysis of
  the NN scattering phase shift \cite{NN-review-2}, 
    the Schr\"{o}dinger equation
  with a certain parametrization of $V_{\rm NN}$ 
   is  solved and compared with the data.
  On the other hand, if we can calculate 
   $\phi({\vec r})$ directly from lattice simulations for various $E$,
   Eq.(\ref{schrodinger.eq2}) can be used to define the 
  non-local potential $U(\vec r,{\vec r}') $ directly without
   recourse to the experimental inputs except for quark masses
    and the QCD scale parameter. 
  In this Letter, instead of finding $U$ by varying $E$, 
   we take only the leading term in the derivative expansion
  at low energies  and extract the central potential $V_{\rm  C}(r)$ 
  at fixed $E$ through   
 \begin{equation}
  V_{\rm  C}(r)
  = E + \frac1{2\mu}
  {
    \vec\nabla^2 \phi(r)
    \over
    \phi(r)
  }.
  \label{CP-PACS-formula}
\end{equation}

 On the lattice, $\phi({\vec r})$ with zero angular momentum
 ($\ell=0$) is defined from the equal-time BS  amplitude as
\begin{eqnarray}
  \phi(\vec r)
  &\equiv&
  \frac1{24}
  \sum_{{\cal R}\in O}
  \frac1{L^3}
  \sum_{\vec x}
  \label{bs.wave.function}
  \\\nonumber
  &\times&
P^{\tau}_{ij} P^{\sigma}_{\alpha\beta}
  \left\langle 0 \left|
  N^i_{\alpha}({\cal R}[\vec r] + \vec x)
  N^j_{\beta} (\vec x)
  \right|
  {\rm NN}
  \right\rangle,
\end{eqnarray}
where we  choose the local interpolating operator for the nucleon: 
 $N^i_{\alpha}
 = \epsilon_{abc} \left( ^{\rm t}q^a C\gamma_5 \tau_2 q^b \right) q^{i,c}_{\alpha}$
  with  $a$, $b$  and $c$ the color indices,
 $\alpha$ and $\beta$ the Dirac indices, $i$ and $j$ the isospin
  indices, and $C\equiv \gamma_4\gamma_2$ the  charge conjugation.
 $\vec r$ describes  the spatial separation  between the nucleons.
  Since we consider the NN scattering at low  energies, we take 
only  the upper components of $N^i_{\alpha}$.
The summation over the vector $\vec x$ projects out 
 the state with zero total-momentum.  
 The summation over  discrete rotation ${\cal R}$ of the 
  cubic group $O$   projects out the $A_1^+$ representation 
 which contains $\ell=0$ state and $\ell \ge 4$ states.
  The former can be singled out by selecting  the 
   lowest energy state with the procedure given 
    in Eq.(\ref{eq:4-point}).
 The spin (isospin) projection is carried out by the operator
  $P^{\sigma}$ ($P^{\tau}$); for example,
  $P^{\sigma}_{\alpha\beta}=(\sigma_2)_{\alpha\beta}
  (=\delta_{\alpha\beta})$
in the spin-singlet (spin-triplet) channel.
 The renormalization factor $Z$, which relates the
  BS amplitude on the lattice and that in the continuum,
 cancels out in  $V_{\rm  C}(r)$.

% Hatsuda version   

  {The $\phi(\vec r =\vec  y - \vec x)$ in Eq.(\ref{bs.wave.function})
   is nothing  but the probability amplitude  to find ``nucleonlike''
   three quarks located at point $\vec x$ and another ``nucleonlike''
   three quarks located  at point $\vec y$.  In  terms of the physical
   states,  $\phi(\vec r)$  contains  not only  the elastic  amplitude
   NN$\rightarrow$NN  but also  the  inelastic amplitudes  such as  NN
   $\rightarrow$ $\pi$NN.   However, at low  energies below threshold,
   the inelastic part  is spatially localized and does  not affect the
   asymptotic  form of  $\phi(\vec r)$.   Note also  that  a different
   choice of the nucleon  interpolating operator modifies the relative
   weight of the elastic and  inelastic amplitudes and thus leads to a
   different   NN  potential.   Nevertheless,   they  give   the  same
   scattering phase shift since  the asymptotic form of $\phi(\vec r)$
   is  independent  of the  interpolating  operators.  It  is an  open
   question at the  moment how the short distant  structure of $V_{\rm
   NN}$  to  be  shown  below   is  affected  by  the  change  of  the
   interpolating  operator.  For  more  details on  these points,  see
   \cite{IAH_full}.  }

% Aoki version  

% {\bf 
%  We note here that  $\phi(\vec r)$ defined above
%  contains not only the elastic part NN$\rightarrow$NN
%  but also inelastic part such as NN $\rightarrow$ $\pi$NN.  
%  {\it A different choice of the interpolating operator would change
%  the relative weight between the elastic and the inelastic parts.
%  Therefore the resultant NN potential may depend on the choice of 
%  interpolating operators, in particular at short distances.
%  At long distances, on the other hand, the resultant NN potential is universal,
%  since, at low energies below threshold, the inelastic contribution is 
%  spatially localized and does not affect the asymptotic behavior
%  (the phase shift) of $\phi(\vec r)$. 
%  In this sense, the NN potential defined from
%  eqs.(\ref{bs.wave.function}) and (\ref{CP-PACS-formula}) is not unique 
%  but is considered as a characteristic representative which
%  gives the universal long distance behaviour.}
%  For more details on these points, see \cite{IAH_full}.  
% }
 
In the actual simulations,
\Eq{bs.wave.function} is obtained
through the four point nucleon correlator,
\begin{eqnarray}
& & \! \! \! \! \! \! \! \! \! \! \! \!
     \! \! \! \! \! \! \! \! \! \! \! \!
      F_{\rm NN}(\vec x,\vec y, t; t_0)
  \equiv
  \left\langle 0 \left|
  N^i_{\alpha}(\vec x,t)
  N^j_{\beta} (\vec y,t)
\overline{\cal J}_{\rm NN}(t_0)
  \right| 0 \right\rangle \nonumber
  \\
  &=&
  \sum_{n}
  A_n\;
  \left\langle 0 \left|
  N^i_{\alpha}(\vec x)
  N^j_{\beta} (\vec y)
  \right| n \right\rangle\;
  e^{-E_n(t - t_0)}.
\label{eq:4-point}
\end{eqnarray}
Here  $\overline{\cal J}_{\rm NN}(t_0)$ is a source term located at $t=t_0$,
which produces the nucleons  with zero total momentum.
  To enhance  the ground
state contribution of the NN system, we  adopt the wall source,
${\cal J}_{\rm NN} (t_0)= P_{ij}^{\tau} P_{\alpha \beta}^{\sigma}
{\cal N}_{\alpha}^i(t_0) {\cal N}_{\beta}^j(t_0)$,
where ${\cal N}$ is obtained from $N$ by replacing
$q$ by $Q(t_0)=\sum_{\vec x} \ q(\vec{x},t_0)$.
$E_n$  is the  energy of the two-nucleon  state $|n\rangle$ and 
 $A_n(t_0) \equiv  \langle n | \overline{\cal J}_{\rm NN}(t_0) |0\rangle$.
  Because of the finite lattice volume $L^3$,
  the energy $E$ takes discrete value and has a finite  shift
   from  the non-interacting case
    $\Delta E = O(1/L^3)$ to be
  determined from the  simulations \cite{luescher}.
%  Note that  $E$ for the scattering states
%  can be negative, if there is attraction in a finite box.

In this Letter,
  we focus  on the spin-singlet and spin-triplet
  channels with zero orbital angular momentum.
 In the standard notation, the former (latter) corresponds to the
  $^{2s+1}\ell_J$=$^1$S$_0$ (=$^3$S$_1$) channel, where
   $s$, $\ell$ and $J$ denote the total spin, orbital angular momentum, and the
    total angular momentum of the two nucleons.
  The $^1$S$_0$ is the simplest channel where only
  the central potential $V_{\rm  C}(r)$ contributes.
  On the other hand, there arises a mixing between  the $^3$S$_1$ and
  $^3{\rm D}_1$ channels because of the tensor force $V_{\rm T}(r)$.
  In this case, one may define an effective central potential
  $V_{\rm C}^{\rm eff}(r)$ which consists of
  the bare central potential and the induced central potential by the 
   $^3{\rm D}_1$ admixture \cite{NN-review-1}. The definition
  in Eq.(\ref{CP-PACS-formula}) with $\phi({\vec r})$
  being projected onto $^1$S$_0$ ($^3$S$_1$)   corresponds  to
   the central potential (the effective central potential).

\begin{figure}[b]
\begin{center}
\includegraphics[width=0.65\figwidth,angle=270]{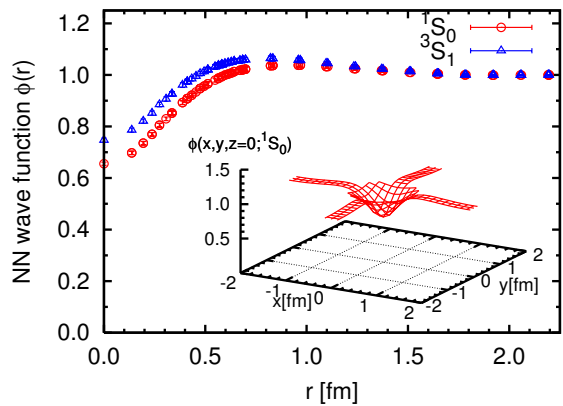}
\end{center}
\caption{The lattice  QCD result of the radial dependence of the
NN wave  function
at $t-t_0=6$ in the $^1$S$_0$ and $^3$S$_1$ channels. Inset shows
  the two-dimensional view in the $x-y$ plane.}
\label{wave-function}
\end{figure}

  To calculate  $\phi({\vec r})$,
  we have carried out simulations
  on  a $32^4$ lattice in the quenched approximation.
  We employ  the 
 plaquette  gauge  action with the gauge  coupling
  $\beta=5.7$  and the Wilson quark action. 
 The lattice spacing determined from
 the $\rho$  meson mass in the chiral limit is
$a^{-1}=1.44(2)$ GeV ($a \simeq 0.137$ fm) \cite{kuramashi},
 which leads to the lattice size $(4.4 \ \mbox{fm})^4$.
 The hopping parameter is chosen to be $\kappa=0.1665$, which
  corresponds to $m_{\pi}\simeq  0.53$ GeV, $m_{\rho}\simeq 0.89$
GeV and  $m_{N}  \simeq  1.34$ GeV.   
We use the global heat-bath algorithm 
with overrelaxations to generate the gauge configurations.
After   skipping  3000 sweeps for   thermalization,  
500 gauge  configurations  are collected with
the interval of  200  sweeps.
 Results for lighter and heavier quark
masses with higher statistics will be reported in \cite{IAH_full}.
The Dirichlet (periodic) boundary condition for quarks
 is imposed  in the temporal (spatial) direction.
To avoid the boundary effect,  the wall source is placed at $t=t_0=5$ at which
the Coulomb gauge fixing is made.
The  ground state  saturation for $t-t_0 \ge 6$ is checked by the
effective mass of the two-nucleon system.

\Fig{wave-function}  shows  the lattice  QCD  result  of  the  wave
function at the time-slice $t-t_0=6$. They are  normalized 
at the spatial boundary $\vec r = (32/2=16,0,0)$.
All the data  including the off-axis ones are  plotted
for $r \alt  0.7$ fm,    beyond  which we  plot  only the  data
locating on the coordinate axes and their nearest neighbors.
 As is clear from \Fig{wave-function},
 the wave function is suppressed at short distance and
 have a slight enhancement at medium distance, which 
  suggests that the NN system has a repulsion (attraction)
  at short (medium) distance.

\begin{figure}[t]
\begin{center}
\includegraphics[width=0.67\figwidth,angle=270]{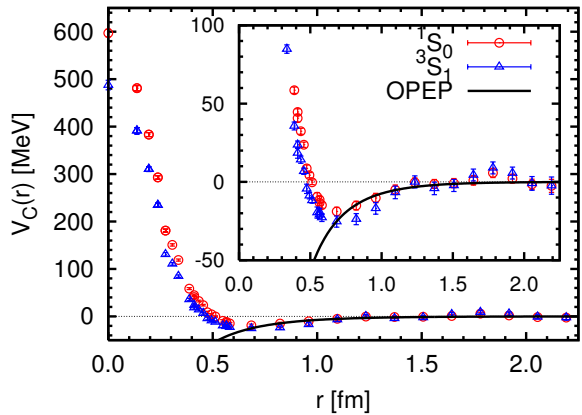}
\end{center}
\caption{The lattice QCD result of the central (effective central)
part of the  NN potential $V_{\rm C}(r)$ ($V_{\rm C}^{\rm eff}(r)$) in the
$^1$S$_0$ ($^3$S$_1$) channel for $m_{\pi}/m_{\rho}=0.595$. The inset shows its
enlargement.
The solid lines correspond to the
  one-pion exchange  potential (OPEP) given in Eq.(\ref{NN-OPEP}).}
\label{NN-potential}
\end{figure}

\begin{figure}[t]
\begin{center}
\includegraphics[width=0.67\figwidth,angle=270]{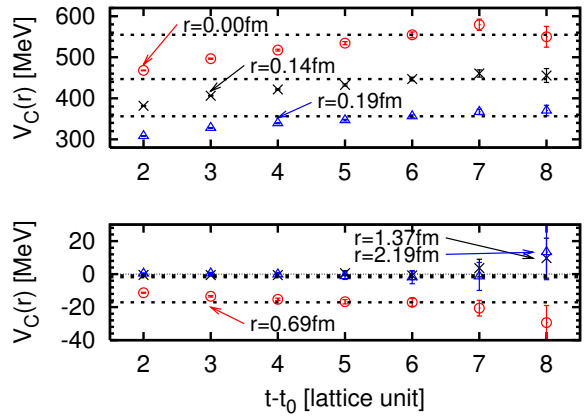}
\end{center}
\caption{$t-t_0$ dependence of $V_{\rm C}(r)$ in the $^1$S$_0$ channel
for several different values of the distance $r$.}
\label{NN-pot-conv}
\end{figure}

\Fig{NN-potential} shows the  central (effective central) NN potential
in the $^1$S$_0$ ($^3$S$_1$)  channel at $t-t_0=6$.  As for $\nabla^2$
in \Eq{CP-PACS-formula},  we take the  discrete form of  the Laplacian
with the  nearest-neighbor points.  $E$  is obtained from  the Green's
function  $G({\vec r};  E)$  {which  is a  solution  of the  Helmholtz
equation  on  the  lattice}  \cite{ishizuka}.   By  fitting  the  wave
function  $\phi({\vec r})$  at the  points ${\vec  r}=(10-16,0,0)$ and
$(10-16,1,0)$  by  $G({\vec  r};  E)$,  we obtain  $E(^1{\rm  S}_0)  =
-0.49(15)$ MeV and $E(^3{\rm S}_1) = -0.67(18)$ MeV.  Namely, there is
a slight attraction between the two nucleons in a finite box.  To make
an independent check  of the ground state saturation,  we plot the $t$
dependence  of $V_{\rm  C}(r)$  in the  $^1$S$_0$  channel at  several
distances   $r=0,  0.14,   0.19,  0.69,   1.37  $   and  2.19   fm  in
\Fig{NN-pot-conv}.  The  saturation indeed  holds  for  $t-t_0 \ge  6$
within errors.

As anticipated from \Fig{wave-function},  $V_{\rm C}(r)$
and $V_{\rm C}^{\rm eff}(r)$ have
repulsive  core at $r \alt 0.5$ fm  with the height of about a few hundred MeV.
Also, they have an attraction of about  $-$(20$-$30) MeV at the
distance $0.5 \alt r \alt 1.0$ fm.
 The solid lines in \Fig{NN-potential}
  show the one pion exchange contribution to the central potential calculated from
\begin{eqnarray}
  V_{\rm C}^{\pi}(r)=
  \frac{g_{\pi N}^2}{4\pi}
    \frac{({\vec \tau}_1\cdot {\vec \tau}_2)({\vec \sigma}_1\cdot {\vec \sigma}_2)}{3}
   \left( \frac{m_{\pi}}{2m_{\rm N}} \right)^2 \frac{e^{-m_\pi r}}{r},
   \label{NN-OPEP}
\end{eqnarray}
 where we have used $m_{\pi}\simeq  0.53$ GeV
   and  $m_{\rm N}  \simeq  1.34$ GeV to be consistent with our data, while 
   the physical value of the $\pi$N coupling constant is used,
    $g_{\pi N}^2/(4\pi)\simeq 14.0$.
  Even in the quenched approximation, the one pion exchange is possible
 as the connected quark exchange between the two nucleons. In
addition, there is in principle a quenched artifact to 
  the NN potential from the flavor-singlet hairpin diagram (the ghost exchange)
 between the nucleons  \cite{savage2}.
Its contribution to the central potential reads \cite{ghost-com}:
%  \begin{eqnarray}
$   V_{\rm C}^{\eta}(r) =   \frac{g_{\eta N}^2}{4\pi}
   \frac{{\vec \sigma}_1\cdot {\vec \sigma}_2}{3}
   \left( \frac{m_{\pi}}{2m_{\rm N}} \right)^2
   \left(\frac{1}{r}
    - \frac{m_0^2}{2m_{\pi}} \right) e^{-m_\pi r}.
$
%  \label{NN-ghost}
%  \end{eqnarray}
  Here $g_{\eta N}$ and $m_0$ are the $\eta$N coupling constant
  and a mass parameter of the ghost, respectively.
 The ghost potential
   has an exponential tail which
  dominates over the Yukawa potential at large distances.
  Its significance can be estimated by 
  comparing the sign and the magnitude of  $e^{m_{\pi}r} V_{\rm C}(r)$
  and $e^{m_{\pi}r} V_{\rm C}^{\rm eff}(r)$ at large distances, because
 $V_{\rm C}^{\eta}(r)$ has an opposite sign between 
 $^1$S$_0$ and $^3$S$_1$.   Our present data shows no evidence of the
  ghost at large distances  within errors,
  which may indicate $g_{\eta N} \ll g_{\pi N}$.
  
Several comments are in order here:\\
\noindent
1.  
%$\phi({\vec r})$  provides us with an 
%   orthodox way to extract the NN scattering length:
  The asymptotic wave function at low energy
 ($E \rightarrow 0$) is approximated as
 $  \phi_{\rm asy}(r) =  { \sin\left(kr + \delta_0(k)\right)
    \over kr } \rightarrow  {r + a_0 \over r}  $, where  
  $\delta_0(k)$ ($a_0$) is the s-wave scattering phase shift (scattering length).
%  and $a_0$
%     \equiv \lim_{k\to 0} \delta_0(k)/k$
%     is the scattering length.
  From the zero of  $r \phi_{\rm asy}(r)$,
   we find $a_{0}(^1{\rm S}_0)= 0.066(22)$ fm and
   $a_{0}(^3{\rm S}_1)= 0.089(27)$ fm  under the assumption 
   that $E \sim -0.5 $ MeV is small enough \cite{luscher-scatt}.
   The reason of having such a small $a_0$ 
   is easily understood from the well-known formula in the Born approximation:
  $a_{0} \simeq - m_{\rm N} \int_0^{\infty}  V_{\rm C}(r) r^2 dr$, where
  (i) the volume factor $r^2 dr$  hides
   the repulsive core at short distances, and (ii) $a_0$ is
   a subtle quantity { which suffers from} a large
   cancellation between repulsion and attraction.\\
\noindent
2. The above points (i) and (ii)  also provide a reason why it is
dangerous at the moment to compare $a_0$
obtained in unphysical quark masses with $a_0$ in experiments
 \cite{savage1}.
   A slight change of the  depth
    and height of the potential
  by the change of the quark mass may affect the
 scattering length substantially.\\
\noindent
3. If the attraction becomes
  large enough at small $m_{\pi}$, 
  the system may  have a bound state and
  $a_0$ changes sign from 
   positive to negative. In the real world, this happens
    in $^3$S$_1$ (the deuteron channel).
%    Note that the above mentioned formula
%   in the Born approximation is not valid any more
%    in such a situation.
 Indeed, the inset of \Fig{NN-potential} shows
that $^3$S$_1$ has stronger attraction than $^1$S$_0$. 
  Note that this difference cannot be
      attributed to the ghost contribution, because it is
     repulsive (attractive) in $^3$S$_1$ ($^1$S$_0$). \\
\noindent
{4.  Our preliminary result with the lighter quark mass 
 at $\kappa=0.1678$ ($m_\pi  \simeq 0.36$ GeV,
  $m_\rho \simeq 0.84$ GeV and  $m_N   \simeq 1.2$ GeV) \cite{IAH_full}
 shows that the height of the  repulsive core increases by 40 \% 
 at the origin, while the minimum of the potential is shifted to 
  $r \sim$ 0.8 fm with approximately the same depth. }

  In  summary, we have  studied  the NN interaction  using
   the lattice QCD simulations in the quenched approximation
   on a (4.4 fm)$^4$ lattice with the quark mass corresponding to
   $m_{\pi}/m_{\rho}=0.595$. 
  We define a NN potential
  with the use of the equal-time Bethe-Salpeter  amplitude
  for two local interpolating operators of the nucleon. 
 The central (effective central) part of the potential
  in the $^1$S$_0$ ($^3$S$_1$) channel
 at low energies  turns out to have a repulsive core
  at short distances surrounded  by attractive well at
  medium and long distances. These properties are  known to be
  the important features of the  phenomenological NN potential.
   The long range tails of $V_{\rm C}(r)$ in both channels are
  consistent with 
the one-pion exchange within statistical errors.
      We did not find numerical evidence of the
  ghost exchange (quenched artifact) for $m_{\pi}/m_{\rho}=0.595$.
 {Preliminary calculation with the lighter quark mass 
($m_{\pi}/m_{\rho}=0.433$)
 shows that the short range repulsion becomes stronger
  and the range of the potential becomes wider.}
  
  It would be quite interesting
  to derive the tensor and  spin-orbit  forces by
  making appropriate projections of the wave function.
  Studies of the  hyperon-nucleon
 and   hyperon-hyperon  potentials, whose
 experimental information is currently very limited,
 are now under investigation: They are 
 particularly important for the physics of {hyper nuclei and
  neutron star core}.
 To unravel the physical origin of the 
 repulsive core, we need further studies on its
  quark mass dependence, channel dependence and 
  interpolating-operator dependence.
  Also, the full QCD simulations at the physical quark mass 
  are necessary to be done in the future.

\vspace{0.2cm}

 The authors thank N. Ishizuka, H. Nemura, 
 D. Kaplan, U. van Kolck, R. Machleidt, S. Sasaki, M. Savage,
 and M. Hjorth-Jensen  for useful comments.
 N.I. thanks N.~Shimizu for useful information and discussions. 
 This research  was supported  in part by the
 Grant-in-Aid of MEXT (Nos. 13135204, 15540251, 15540254, 18540253).
 Our simulations have been performed with IBM Blue
 Gene/L at KEK under a support of its Large Scale Simulation Program, No.18 and
 No.06-21 (FY2006).


\begin{references}
\bibitem{ghost-com}
With another ghost parameter
$\alpha_{\Phi}$, the formula is modified as $1/r \rightarrow (1-\alpha_{\Phi})/r$
and $m_0^2 \rightarrow m_0^2 - \alpha_{\Phi}m_{\pi}^2$.
\bibitem{luscher-scatt}
 The L\"{u}scher's  finite volume formula \cite{luescher}
 together with $E$ obtained in the text 
 lead to $a_{0}(^1{\rm S}_0)=0.123(39)$ fm and
  $a_{0}(^3{\rm S}_1)=0.17(5)$ fm, which  are consistent with those
   obtained from the wave function.
\end{references}
\end{document}